\documentclass{article}
\usepackage{epsfig}
\newcommand{\bfr}{\begin{flushright}}
\newcommand{\efr}{\end{flushright}}
 
\begin{document}
\title{Cosmic Strings in Compacitified Gauge Theory
}
\author{ 
Atsushi Nakamula and Satoru Hirenzaki\\
Department of Physics, Tokyo Metropolitan University,\\
Setagaya-ku, Tokyo 158, Japan\\
and\\
Kiyoshi Shiraishi\\
Institute for Nuclear Study, University of Tokyo, \\
Midori-cho, Tanashi,
Tokyo 188, Japan
}
\date{Nuclear Physics {\bf B339} (1990) pp. 533--544 
}
\maketitle
\begin{abstract}
A solution of the vortex type is given in a six-dimensional
$SU(2)\times U(1)$ pure gauge theory coupled to Einstein gravity in a
compactified background geometry. We construct the solution of an
effective abelian Higgs model in terms of dimensional reduction. The
solution, however, has a peculiarity in its physically relevant
quantity, a deficit angle, which is given as a function of the ratio of
the gauge couplings of $SU(2)$ and $U(1)$. The size of the extra space
(sphere) is shown to vary with the distance from the axis of the
``string''. 
\end{abstract}

In the course of
universe evolution, phase transitions are a very important subject in
cosmology and particle theory \cite{1}. The most attractive features are
the generation of topological defects \cite{2} such as domain walls,
cosmic strings, and so on. Especially cosmic strings provide a mechanism
to generate seeds of galaxies, and thus their cosmological roles are of
great significance \cite{2,3}. 

There are two types of cosmic strings,
namely global and local strings, which are distinguished by the type of
symmetry broken via the associated phase transition. The global strings
are formed when global symmetries break down such as in $U(1)$ scalar
theory, whereas the local strings are subject to gauge symmetry
breaking. 

The simplest approach to the study of local cosmic strings in
ordinary threedimensional space is to examine the abelian Higgs vortex,
i.e. the Nielsen-Olesen type solution \cite{4,5} to equations of motion
for the scalar and $U(1)$ gauge fields. This solution represents an
infinitely long isolated string in flat space-time. If one wants to
consider a vortex or a string in a cosmological situation more
precisely, however, we must couple the system with gravity since the
core of the string has finite energy density. 

In this context, Garfinkle
\cite{6} found consistent solutions of the Einstein-scalar-$U(1)$ gauge
system with an exact energy-momentum of the matter fields in the
situation that the vacuum expectation value of the scalar field is much
smaller than
 the Planck scale. The
characteristic feature of these string solutions is that at radial
infinity, the space-time manifold becomes Minkowski space-time minus a
wedge, i.e. a deficit angle appears around the cosmic string. Moreover,
Laguna-Castillo and Matzner \cite{7} solved the same system numerically
without any assumption except for boundary conditions. They showed that
when the symmetry breaking scale, or the vacuum expectation value of the
scalar field, is greater than $10^2$ GeV, the curvature of the space-time
manifold is not negligible. 

In more realistic models, the abelian gauge
symmetry might come from some non-abelian symmetry. If we believe in a
grand unified gauge symmetry, the $U(1)$ gauge symmetry of the abelian
Higgs model is contained in the large group. 

On the other hand, from the
perspective of unified theories including gravity, higher-dimensional
theories have very attractive possibilities \cite{8}. For instance the
string theory \cite{9} is naturally formulated in more than four
dimensions and is regarded as a candidate for a finite theory of
gravity. In these theories we must assume that the extra space has
become a very tiny internal space which is sufficiently small never to
be seen by present experiments. Nevertheless, it is very interesting to
speculate on the effects such extra dimensions might have on the
four-dimensional world we live in. For instance, it is permissible that
the vector fields on the internal space have non-vanishing vacuum
expectation values, which do not break Poincar\`e invariance in the
four-dimensional world. Therefore, gauge symmetry breaking can be caused
by the vacuum gauge fields \cite{10}. 

The effects of the vacuum
configuration on the extra space may also play an essential role as the
source of spontaneous compactification, together with the cosmological
constant. In such a case the gauge field has non-zero field strength on
the compact space with curvature \cite{8,11}. 

If ``elementary''
non-abelian gauge fields exist in such higher-dimensional theories,
another interesting possibility comes about. It was shown that gauge
boson-Higgs scalar systems are derived from dimensional reduction of
higherdimensional Yang-Mills theory \cite{12}. Many authors considered
dimensional reduction as a device to obtain various breaking patterns of
the Higgs mechanism. We wish to make use of this scheme to investigate
the multidimensional universe. 

In this paper, we consider the full
coupled equations of motion of effective four space-time dimensional
Einstein-scalar-$U(1)$ gauge system induced from higherdimensional
Einstein-Yang-Mills-Maxwell theory. Here we take Sz, a sphere, as the
extra space and $SU(2)$ as the non-abelian gauge group. Effective abelian
Higgs-like equations are obtained from the pure Yang-Mills theory. We
solve the solution of the vortex-type numerically. The conclusion is a
very natural one in the sense that the properties of the ``cosmic
string'' are almost the same as those of the ordinary string in
three-dimensional space except that the size of the extra $S^2$ space
varies with the radial coordinate. Moreover in our string,
cosmologically relevant quantities such as the deficit angle and the
scale of the extra space at the core are determined by the ratio between
the $U(1)$ and $SU(2)$ couplings. 

We begin with the following action: 
\begin{equation}
S=\int d^6x
\sqrt{-g}\left(-\frac{1}{2\kappa^2}R+\frac{1}{4e^2}{\rm
tr}(\mathbf{F}_{MN}\mathbf{F}^{MN}
)+\frac{1}{4g^2}G_{MN}G^{MN}+\Lambda\right)\,.
\end{equation}
Here $\mathbf{F}_{MN}$ and $G_{MN}$ are the field strengths of $SU(2)$
and
$U(1)$ gauge fields, respectively. $e$ is an $SU(2)$ gauge coupling
constant while
$g$ is a $U(1)$ gauge coupling; $\Lambda$ is a cosmological constant.
The scalar curvature of SN with unit radius is defined as $R = +N(N -
1)$. The indices $M$ and $N$ take values in six dimensions. Although
these gauge groups can be seen as subgroups of some larger gauge group,
we do not pursue this line in this paper. 

The field equations are
directly derived from this action, 
\begin{eqnarray}
R_{MN}&=&\frac{1}{2}\kappa^2\Lambda g_{MN}+\kappa^2\left(
T_{MN}-\frac{1}{4}T g_{MN}\right)\,,\\
D_M\mathbf{F}^{MN}&=&\nabla_M\mathbf{F}^{MN}+i[\mathbf{A}_M,
\mathbf{F}^{MN}]=0\,,\label{3}\\
\nabla_M G^{MN}&=&0\,,
\end{eqnarray}
where the field strengths are given by
\begin{eqnarray}
\mathbf{F}_{MN}&=&\partial_M\mathbf{A}_N-\partial_N\mathbf{A}_M
+i[\mathbf{A}_M,\mathbf{A}_N]\quad\mbox{for }SU(2)\,,\\
G_{MN}&=&\partial_M B_N-\partial_NB_M\quad\mbox{for }U(1)\,. 
\end{eqnarray}
$\mathbf{A}_M$ and $B_M$ are the $SU(2)$ and $U(1)$
gauge field respectively, and $\nabla_M$ is the covariant derivative.
$T_{MN}$ is the energy-momentum tensor of the gauge fields given as
follows:
\begin{eqnarray}
T_{MN}&=&\frac{1}{e^2}{\rm tr}\left(\mathbf{F}_{MP}\mathbf{F}_N{}^P-
\frac{1}{4}\mathbf{F}_{PQ}\mathbf{F}^{PQ}g_{MN}\right)\nonumber \\
&
&+\frac{1}{g^2}\left(G_{MP}G_N{}^P-\frac{1}{4}G_{PQ}G^{PQ}g_{MN}\right)\,,
\\
T&=&T_{MN}g^{MN}\,.
\end{eqnarray}

For simplicity, the background space-time is assumed to be locally
$M^4\times S^2$, namely the direct product of a flat ``universe'' in
which the cosmic string is embedded and the internal two-dimensional
sphere. Furthermore, we look for static solutions in the present work.
Accordingly, the square of the line element can be set to be
\begin{equation}
ds^2=e^A(-dt^2+dz^2)+dr^2+e^C d\psi^2+e^B d\Omega^2(S^2)\,,
\end{equation}
where $d\Omega^2(S^2)$ represents the line element of the sphere with
unit radius, i.e. $d\theta^2+\sin^2\theta d\phi^2$ and $0\le\psi<2\pi$,
$0\le\theta\le\pi$ and $0\le\phi<2\pi$. Here $e^{A/2}$ and $e^{B/2}$ are
the scale factors of two-dimensional ``homogeneous'' space-time and the
internal sphere, respectively. The axis of the string is located at $r=
0$. The functions $A$, $B$, and $C$ are assumed to depend only on the $r$
coordinate, i.e. on the distance from the axis of the string.

First we assume the
$U(1)$ gauge field being a monopole configuration (with minimal magnetic
charge)
\cite{11}:
\begin{equation}
B_\phi=\frac{1}{2}(\cos\theta\pm 1)\,,
\end{equation}
and the other components of the $U(1)$ vector field are assumed to
vanish. This $U(1)$ monopole is required only to support the large
dimensions by cooperating with the fine-tuned cosmological constant.
Thus we require a relation among the couplings \cite{11}: 
\begin{equation}
\Lambda=2g^2/\kappa^4\,.
\label{11}
\end{equation}

The vacuum configurations of $SU(2)$ gauge fields are crucial to generate
the cosmic string. In our model, we set the components of $SU(2)$ gauge
field on the internal space as
\begin{eqnarray}
\mathbf{A}_\theta&=&\Phi_1\frac{1}{2}
\left(
\begin{array}{cc}
0 & -i e^{-i\phi}\\
i e^{i\phi} & 0
\end{array}
\right)+
\Phi_2\frac{1}{2}
\left(
\begin{array}{cc}
0 &  e^{-i\phi}\\
 e^{i\phi} & 0
\end{array}
\right)\,,\\
A_\phi&=&-\Phi_1\frac{1}{2}
\left(
\begin{array}{cc}
0 &  e^{-i\phi}\\
 e^{i\phi} & 0
\end{array}
\right)\sin\theta+
\Phi_2\frac{1}{2}
\left(
\begin{array}{cc}
0 & -i e^{-i\phi}\\
i e^{i\phi} & 0
\end{array}
\right)\sin\theta\,,\nonumber \\
& &+
\frac{1}{2}
\left(
\begin{array}{cc}
1-\cos\theta & 0\\
0 & -(1-\cos\theta)
\end{array}
\right)\,.
\end{eqnarray}

Here $\Phi_1$ and $\Phi_2$ are the functions independent of the $S^2$
coordinates $\theta$ and $\phi$. From the four-dimensional point of
view, they can be seen effectively as components of a complex scalar
field (see eqs. (14a, b)). Note that when $\Phi_1=\Phi_2=0$ this gauge
configuration is a magnetic monopole on $S^2$; therefore, it has a finite
energy density. Further, the large-dimensional components of the $SU(2)$
gauge field are assumed to have the following form:
\begin{equation}
\mathbf{A}_\mu=A_\mu(x^\mu)\frac{1}{2}\left(
\begin{array}{cc}
1 & 0\\
0 & -1
\end{array}
\right)\,,
\label{13}
\end{equation}
where $A_\mu(x^\mu)$ depends only on the coordinates of the four
space-time dimensions but not on the extra dimensions. $\mu$ runs over
the four coordinates. Note that here we use a coordinate basis associated
with the metric and not an orthonormal one. 

We emphasize that these
components of the gauge field play the roles of the ``$U(1)$'' gauge
field in four dimensions in our scheme as we will show in the following.
Note that, of course, this effective ``$U(1)$'' fields has no relation
to the six-dimensional $U(1)$ field introduced earlier. 

With respect to
these classical gauge configurations, the equations of motion (\ref{3})
are reduced as follows:
\begin{eqnarray}
D^\mu D_\mu\hat{\Phi}+\frac{1}{e^B}(1-|\hat{\Phi}|^2)\hat{\Phi}&=&0\,,\\
\nabla_\mu(e^B F^{\mu\nu})+i(\hat{\Phi}^*D^\nu\hat{\Phi}-
\hat{\Phi}D^\nu\hat{\Phi}^*)&=&0\,,
\label{14}
\end{eqnarray}
where $\hat{\Phi}=\Phi_1+1\Phi_2$ and $F_{\mu\nu}=\nabla_\mu A_\nu-
\nabla_\nu A_\mu$. The covariant derivative is defined as
$D_\mu=\nabla_\mu+iA_\mu$ with $\nabla_\mu$ being the covariant
derivative associated with the four-dimensional metric $g_{\mu\nu}$.

These equations closely resemble those of the abelian Higgs model, which
has in general two couplings between ``matter'' and gauge fields. In our
model, the ``scalar'' self-interaction and ``gauge'' coupling constants,
which are implicitly contained in eq.~(\ref{14}) in our notation,
originate from the same Yang-Mills coupling. Another different point is
that the effective equations contain the extra dependence on the
position through the factor $e^B$. 

Before solving these equations coupled
with Einstein equations, we must give an ansatz for the gauge field in
order to generate an isolated string. This requires
\begin{eqnarray}
\hat{\Phi}&=&X(r)e^{i\psi}\,,\\
A_{\psi}(r)&=&P(r)-1\,,\quad\mbox{otherwise } A_\mu(r)=0\,.
\end{eqnarray}
As is well known, the solution of this type has unit winding number around the axis
\cite{4,5}.

With all these assumptions, the Yang-Mills equation (\ref{3}), which is
equivalent to eq. (\ref{14}), is reduced to
\begin{eqnarray}
(KX')'&=&KB'X'+X[K(X^2-1)e^{-2B}+e^{2A+2B}P^2/K]\,,
\label{16a}\\
(e^{2A}P'/K)'&=&e^{2A}[-2B'P'+2X^2 e^{-B}P]/K\,,
\label{16b}
\end{eqnarray}
where $K=\exp(A+B+C/2)$, and the prime means derivative with respect
to $r$. For convenience, we adjust the unit of the length scale such that
$\kappa^2/(4g^2)=1$. In this unit, we find that the asymptotic value of
the radius of the extra sphere at large distance from the center of the
string is equal to one.

With the relation (\ref{11}), the Einstein equations for the variables
$A$, $B$, and $K$ are given in our unit by
\begin{eqnarray}
A''&=&-(K'/K)A'-\frac{1}{2}(1-e^{-2B})+q^{-2}\{e^{-B}{X'}^2+e^{2A+B}
P^2X^2/K^2
\label{17a}\nonumber \\
& &+e^{2A+2B}{P'}^2/(2K^2)+e^{-2B}(X^2-1)^2/2\}\,, \\
B''&=&-(K'/K)B'+2e^{-B}-\frac{1}{2}(1+3e^{-2B})-q^{-2}\{e^{-B}{X'}^2
+e^{2A+B}P^2X^2/K^2
\label{17b}\nonumber \\
& &-e^{2A+2B}{P'}^2/(2K^2)+3e^{-2B}(X^2 -1)^2/2\}\,,\\
K''&=&K(2e^{-B}-\frac{5}{4}-\frac{3}{4}e^{-2B})-q^{-2}\frac{1}{2}K\{
-e^{-B}{X'}^2+3e^{2A+B}P^2X^2/K^2\nonumber \\
& &-e^{2A+2B}{P'}^2/(2K^2)+3e^{-2B}(X^2-1)^2/2\}\,,
\label{17c}
\end{eqnarray}
where $q^2=e^2/(2g^2)$.

To find solutions of these differential equations, we use a numerical method
with the physically acceptable boundary conditions of ref.~\cite{7} and
the conditions on $B$, i.e. the scale of the internal space.

First, at the boundary the behaviors of the ``scalar'' $\Phi$ and the
``$U(1)$ gauge field'' are given by
\begin{eqnarray}
& &X(0)=0\,,\qquad P(0)=1\,,\\
& &X(\infty)=1\,,\qquad P(\infty)=0\,.
\end{eqnarray}
These come from the requirement that there are no singularities of gauge fields
introduced originally at the core of the string and that integrating the
energy-momentum tensor must yield a finite value.

Second, we give the condition for the metric of the four-dimensional part such
that the space-time is locally Minkowski up to a constant, which can be scaled at
the center of the string,
\begin{eqnarray}
& &A(0)=K(0)=0\,,\qquad A'(0)=B'(0)=0\,,
\label{19a}\\
& &B(\infty)=0\,,\qquad K'(0)=e^{B(0)}\,.
\label{19b}
\end{eqnarray}
Eq.~(\ref{19b}) is due to the choice of the unit of the scale,
$\kappa^2/(4g^2)=\kappa^2q^2/(2e^2)=1$.

After fixing all of these boundary conditions, we have solved the full set of
coupled equations (\ref{16a}), (\ref{16b}) and (\ref{17a}-\ref{17c}) by
using a computational program named COLSYS \cite{13}. We determine
$B(0)$ by a guess to converge the solution iteratively. 

\begin{figure}[ht]
\begin{center}
\includegraphics[width=5cm]{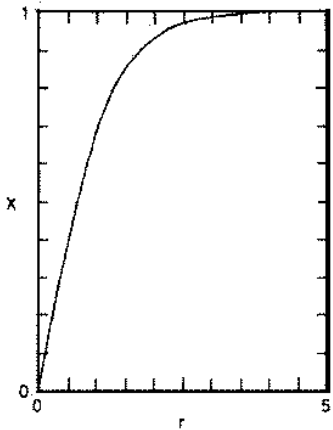}
\caption{Solution for the effective scalar field $X$ for $q=5$. The
curve is hardly sensitive to the value of $q$ ($\ge 2$).}
\label{f1}
\end{center}
\end{figure}

\begin{figure}[ht]
\begin{center}
\includegraphics[width=5cm]{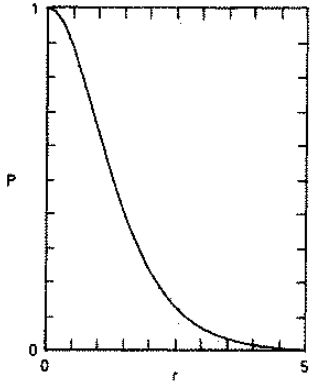}
\caption{Solution for the effective gauge field $P$ for $q=5$. The
curve is hardly sensitive to the value of $q$ ($\ge 2$).}
\label{f2}
\end{center}
\end{figure}

The numerical
results are shown in figures. The behavior of the effective abelian
Higgs fields indicates the characteristic configuration (figs.~\ref{f1}
and \ref{f2}), which is very similar to the ordinary model (in
four-dimensional space-time). The width of the string core in our model
is $O(1)$ in our unit, i.e. the same order as the size of the extra
space at radial infinity.

\begin{figure}[ht]
\begin{center}
\includegraphics[width=5cm]{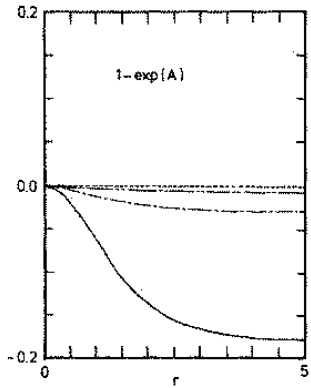}
\caption{Solution for the metric field $1-e^A$ as a function of the
radial coordinate $r$. The lines (---------), (--- {-} ---),  (--- - -
---), and (- - - - - -) correspond to the parameter $q =2, 5, 10$, and
$100$, respectively.}
\label{f3}
\end{center}
\end{figure}

From fig.~\ref{f3} we can see a new feature of the string in our model;
there exists an attractive gravitational force along the radial
direction in our solution with any value of $q^2$, in contrast to the
case of the ordinary abelian Higgs vortex where a repulsive force is also
allowed depending on the ratio of gauge and scalar self-couplings. This
fact suggests that a single vortex with higher windings of the
``scalar'' and ``gauge'' configuration is stable in a static
circumference as in the case of the vortex in a ``type I
superconductor'' \cite{5}. 

\begin{figure}[ht]
\begin{center}
\includegraphics[width=5cm]{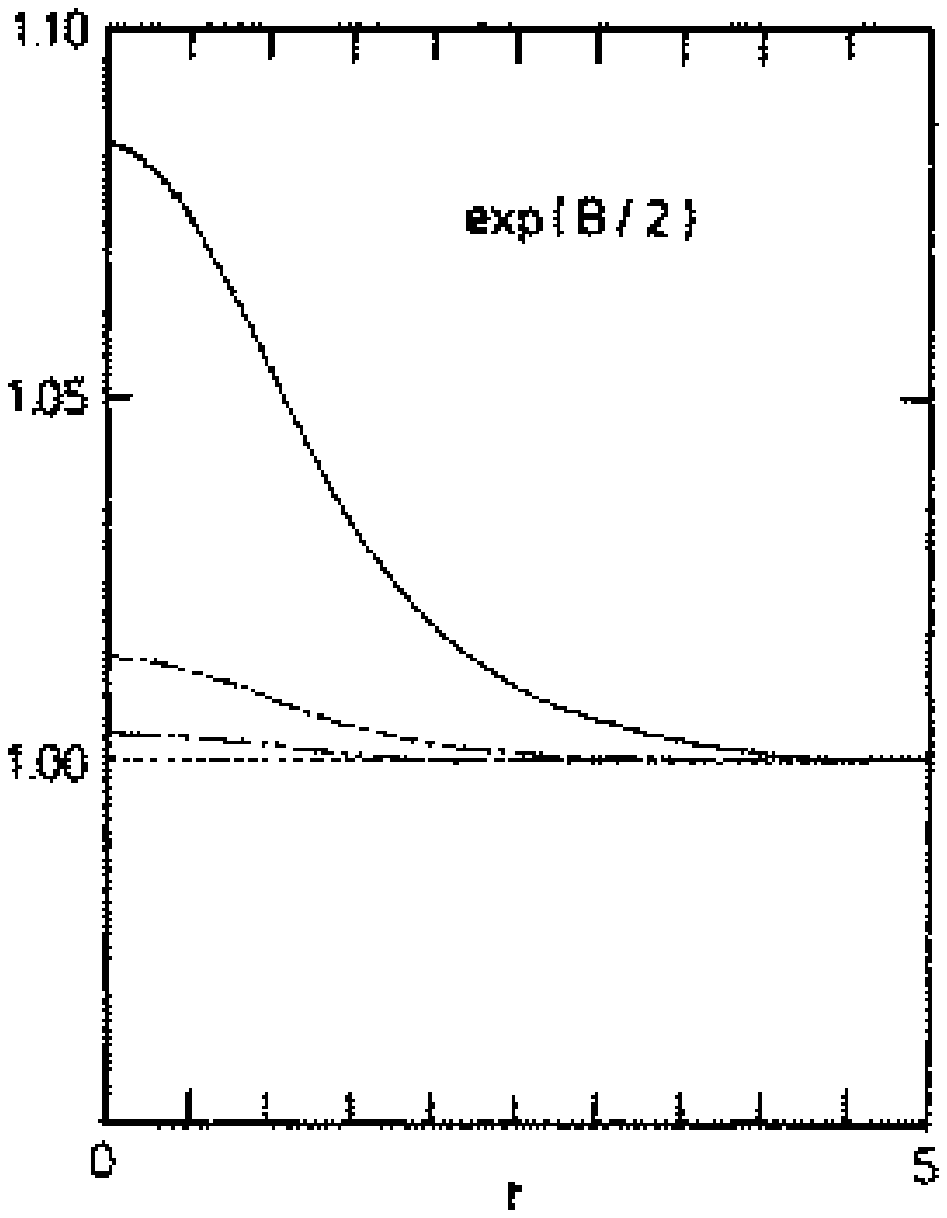}
\caption{The metric component $e^{B/2}$, i.e. the size of the extra space
at the coordinate origin as a function of the radial coordinate $r$. The
line definitions are the same as in fig.~3.}
\label{f4}
\end{center}
\end{figure}
\begin{figure}[ht]
\begin{center}
\includegraphics[width=5cm]{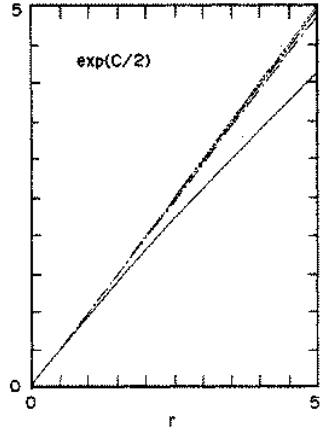}
\caption{Solution for the $e^{C/2}$
metric component as a function of the radial coordinate $r$. The line
definitions are the same as in fig.~3.}
\label{f5}
\end{center}
\end{figure}
\begin{figure}[ht]
\begin{center}
\includegraphics[width=5cm]{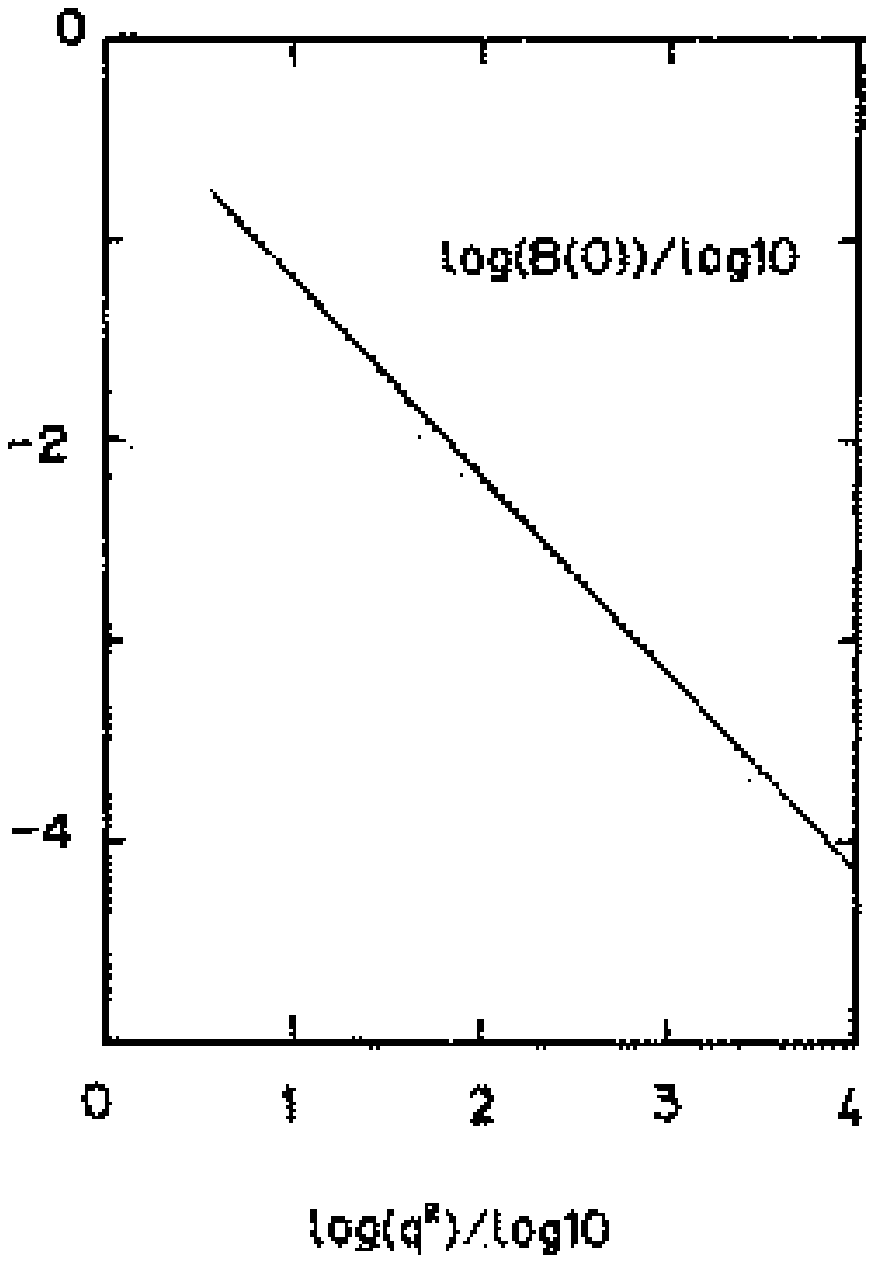}
\caption{Plot of $B(0)$ as a function of $q^2$. This line is well
approximated by $B(0) =q^{-2}\times$(const. of $O(1)$).}
\label{f6}
\end{center}
\end{figure}

Next we consider the deficit angle at radial
infinity. A definition of such a quantity is given in refs. \cite{6,7},
\begin{equation}
\Delta\psi\equiv 2\pi\left\{1-\lim_{r\rightarrow\infty}
\frac{d}{dr}(e^{C/2})\right\}\,. 
\end{equation}

\begin{figure}[ht]
\begin{center}
\includegraphics[width=5cm]{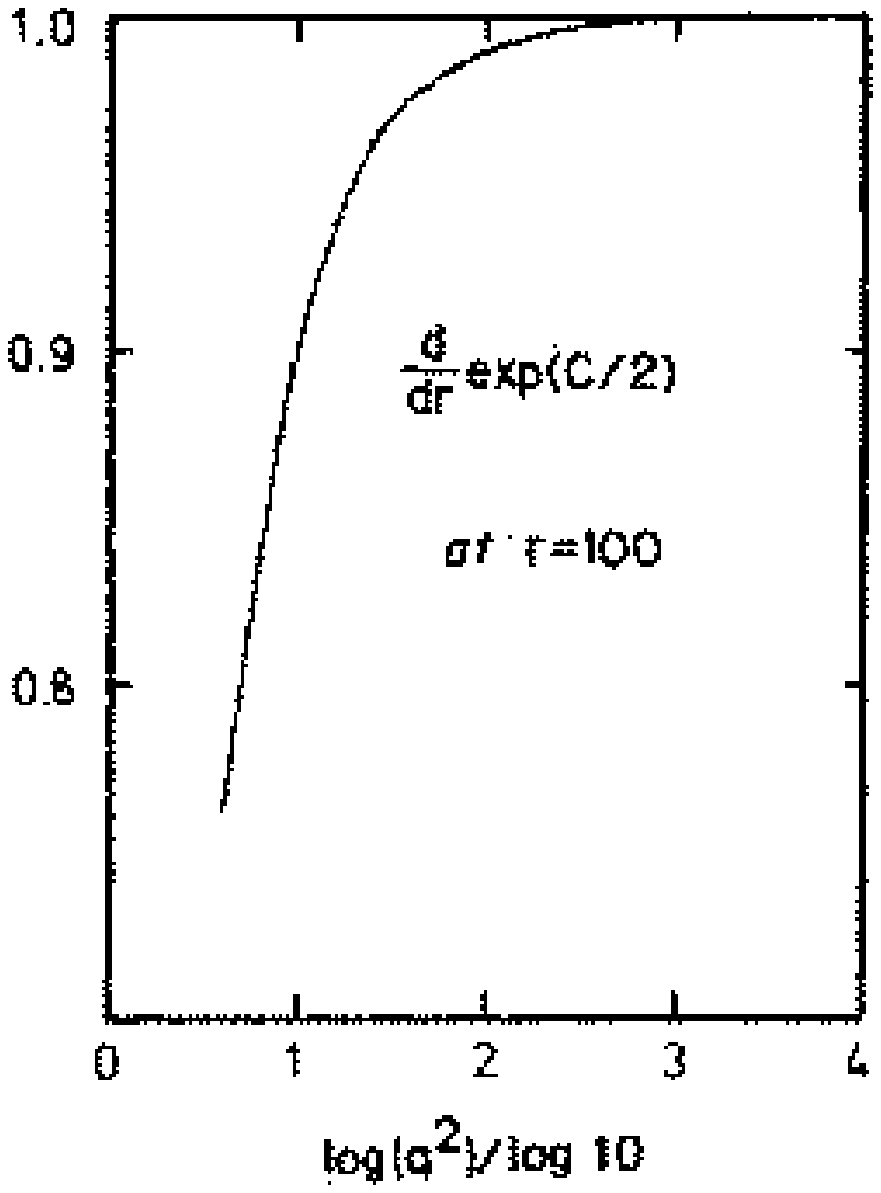}
\caption{Plot of $d/dr(e^{C/2})$ at radial infinity (actually evaluated
at $r=100$) as a function of $q^2$. This curve is well approximated by
$d/dr(e^{C/2})|_{r=100}=1-q^{-2}$.}
\label{f7}
\end{center}
\end{figure}

Our numerical evaluation is shown in fig.~\ref{f7}, and we find that the
deficit angle monotonically decreases with $q^2$. This result is quite
natural. When $q^2$ is large, the energy-momentum of the $SU(2)$ gauge
field is nearly decoupled from the other part in the equations
(\ref{17a}-\ref{17c}), and thus the four-dimensional space-time becomes
flat everywhere. The stability of the structure of the solution is
guaranteed also by the mild behavior of the metric at large $q^2$
(figs.~\ref{f3}, \ref{f4}, and \ref{f5}). The deficit angle turns out to
be well approximated by $\Delta\psi=2\pi q^{-2}$ in the region of the
parameter
$q^2$ where we investigate the solution. 

The property of the internal
space are shown in figs.~\ref{f5} and \ref{f6}. We should pay attention
to the fact that the value of
$\exp(B(0)/2)$, that is, the radius of the internal sphere at the axis
of the string, grows as $q^2$ gets smaller. This radius is
always larger than the value at radial infinity. We find a linear relation in a plot of
$\log B(0)$ versus $\log q^2$. $B(0)$ is expressed approximately as $B(0)
\sim q^{-2} \times (\mbox{constant of } O(1))$. The effect of the
vortex-like configuration of $SU(2)$ gauge field might open a ``window''
to the extra dimensions in the core of the cosmic string, when $q^2\sim
O(1)$.

In ref.~\cite{7}, the parameter $\eta$ indicates the symmetry-breaking
scale in the Planck unit. In our model, the coupling between
energy-momentum of matter fields and gravity is proportional to $q^{-2}$,
instead of $\eta^2$ in the ordinary string model. Roughly speaking, $q^2$
gives the ratio of the radius of the extra space to the Planck length in
the asymptotic region $r\rightarrow\infty$, provided that the Yang-Mills
coupling in fourdimensions is of order $1$.

To summarize, we have obtained a vortex-type solution in
a higher-dimensional gauge theory by a numerical method. The physical
properties of the solution, such as the deficit angle around the string
axis and variation of the size of the extra space has been shown to be
dependent on the ratio of the two couplings in the theory.

There are many subjects for future investigations. One could investigate
the following topics: existence (or non-existence) of the fermionic zero
mode and construction of a superconducting string (possibly from a large
gauge group), phase transition and string formation. In particular the
investigation of the energy spectrum of fermions coupled with the string
background is of great interest from the viewpoint of higher-dimensional
theory. Probably different aspects from ordinary strings will be found
in the study of these problems. Moreover, through this study the
cosmological significance of cosmic strings in higher-dimensional theory
can be argued in detail.

The energy-momentum tensor in our model and cosmic strings in de Sitter
and time-dependent (higher-dimensional) background will be analyzed in a
forthcoming publication. We are also interested in the behavior of the
size of the extra space when a loop of the string is made and collapses,
and when two strings collide or intersect. Such dynamical processes
might provide the possibility of observing the extra dimensions.

Finally, we give a few comments on the case of a four-dimensional
homogeneous spacetime, in which $e^{A/2}$ is the scale factor. We checked
the behavior of all fields at the same value of $q^2$ and they are almost
the same as in the case already mentioned, as expected. The reason why
we wish to consider the exotic case is that
we can investigate the scenario of dimensional reduction, which is
expressed by the slogan, ``we live in a topological defect in a
higher-dimensional space'' \cite{14}. The possibility of this scenario in
an extended version of our model will be reported elsewhere.

\bigskip

The authors would like to thank S.~Saito for reading this manuscript and
for useful comments.

This work is supported in part by the Grant-in-Aid for Encouragement of
Young Scientist from the Ministry of Education, Science and Culture (\#
63790150). 

One of the authors (K.S.) is grateful to the Japan Society
for the Promotion of Science for the fellowship. He also thanks Iwanami
F\=ujukai for financial aid.


\end{document}